# Beliefs and Cooperation

Bernardo A. Huberman and Natalie S. Glance

Dynamics of Computation Group
Xerox Palo Alto Research Center
Palo Alto, CA 94304

**Abstract**

Individuals in groups must often choose between acting selfishly and cooperating for the common good. The choices they make are based on their beliefs on how they expect their actions to affect others. We show that for a broad set of beliefs and group characteristics cooperation can appear spontaneously in non-cooperative groups after very long periods of time. When delays in information are unavoidable the group dynamics acquires a wide repertoire of behaviors, ranging from opportunistic oscillations to bursty chaos, thus excluding the possibility of sustained cooperation.

## 1. Introduction

Social dilemmas have long attracted the attention of sociologists, economists and political scientists because they are central to issues that range from securing ongoing cooperation in volunteer organizations, such as unions and environmental groups, to the possibility of having a workable society without a government. Environmental pollution, nuclear arms proliferation, population explosion, conservation of electricity and fuel, and giving to charity are a few more examples of situations where an individual benefits by not contributing to the common cause, but if all individuals shirk, everyone is worse off. Although there are no simple solutions to social dilemmas, studying them sheds light on the nature of interactions among people and the emergence of social compacts. And because such dilemmas involve the interplay between individual actions and global behavior, they elucidate how the actions of a group of individuals making personal choices gives rise to social phenomena.

Discovering the global behavior of a large system of many individual parts, such as the dynamics of social dilemmas, calls for a bottom-up approach. Aggregate behavior stems from the actions of individuals who act to maximize their utility on the basis of uncertain and possibly delayed information. As long as one is careful both in constructing the model and in making clear its limitations and its underlying assumptions, the insights obtained through such an approach can be very valuable.

Beliefs and expectations are at the core of human choices and preferences. They arise from the intentional nature of people and reflect the way decsion-makers convolve the future as well as the past into decisions that are made in the present. For example, individuals acting within the context of a larger group may take into account the effect of their actions both on personal welfare and on the welfare of the larger group. In other words, individuals form their own models of how the group dynamics works based on some set of beliefs that color their preferences.

In order to better model the dynamics of social dilemmas, we extend our previous framework of individual expectations [1] to one that can accommodate a wider range of beliefs. We describe, in particular, two classes of beliefs, bandwagon expectations and opportunistic expectations. Our framework of expectations rests on two pillars: the first is that agents believe the influence of their actions to decrease with the size of the group; the second is that individuals believe their actions influence others to a degree that depends on how many already contribute to the common good. How agents believe their degree of influence to vary is left unspecified within the broader framework and is instead instantiated through different classes of expectations. Agents with bandwagon expectations believe that their actions will be imitated by others to a greater extent when the overall state tends towards cooperation and to a lesser extent when it tends towards defection. For this class of expectations, agents believe that cooperation encourages cooperation and that defection encourages defection at a rate which depends linearly on the proportion of the group cooperating. Bandwagon agents are thus more likely to cooperate the greater the observed level of cooperation. Opportunistic expectations are similar to bandwagon expectations when the proportion cooperating is small. However, when the proportion is large, opportunistic agents believe that the rate at which cooperation encourages cooperation and defection encourages defection decreases linearly with the fraction cooperating. As a result, agents with opportunistic expectations will "free ride" in a mostly cooperating group since they expect that a small amount of free riding will go unnoticed.

The concept of beliefs and expectations, of course, covers a much broader spectrum than the narrow domain of how one's actions might affect another's. Cultural beliefs concerning "good" and "bad," social norms, personal morality, life-style, external pressure, and many other biases enter into an individual's preferences. These variations among individuals we model in a general way using the notion of diversity [2]. We found in earlier work that in scenarios where diversity can be modelled as a spread about a common set of beliefs then it effectively acts as additional source of uncertainty. Having merged together the various facets of diversity in this fashion, we can now concentrate on how different sorts of expectations affect group dynamics.

For the above two classes of expectations, we show that there is a critical group size beyond which cooperation is no longer sustainable and that below this critical size there is a regime in which there are two fixed points. The stability of these fixed points and their dynamic characteristics are very similar for both classes of beliefs when there are at most small delays in information but differs when the delays become large. For bandwagon expectations, the two fixed points are stable, and in the presence of uncertain information, large-scale fluctuations eventually take the group over from the metastable fixed point to the optimal one. These sudden transitions are both unpredictable and sudden and take place over a time scale that grows exponentially with the size of the group. The same kind of behavior is observed in groups with opportunistic expectations. However, in this case delays in information can play an important role. When the delays are long enough, the more cooperative equilibrium becomes unstable to oscillations and chaos.

Section 2 introduces social dilemmas and their game-theoretical representation and Section 3 develops the theory of their dynamics including expectations. In Section 4, analytical techniques are used to solve for the dynamics: a set of general results covers both types of expectation. Section 5 presents the results of computer simulations which further elucidate the behavior of the system for the two classes of bandwagon and opportunistic expectations. In particular, the computer experiments show how a group can suddenly and unexpectedly transition from one equilibrium to another and how the dynamics of an opportunistic group with delayed information can exhibit oscillations and bursty chaos.

## 2. Social Dilemmas

There is a long history of interest in collective action problems in political science, sociology, and economics [3, 4]. Hardin coined the phrase “the tragedy of the commons” to reflect the fate of the human species if it fails to successfully resolve the social dilemma of limiting population growth [5]. Furthermore, Olson argued that the logic of collective action implies that only small groups can successfully provide themselves with a common good [6]. Others, from Smith [7] to Taylor [8, 9], have taken the problem of social dilemmas as central to the justification of the existence of the state. In economics and sociology, the study of social dilemmas sheds light on, for example, the adoption of new technologies [10] and the mobilization of political movements [11].

In a general social dilemma, a group of people attempts to obtain a common good in the absence of central authority. The dilemma can be represented using game theory. Each individual has two choices: either to contribute to the common good, or to shirk and free ride on the work of others. The payoffs are structured so that the incentives in the game mirror those present in social dilemmas. All individuals share equally in the common good, regardless of their actions. However, each person that cooperates increases the amount of the common good by a fixed amount, but receives only a fraction of that amount in return. Since the cost of cooperating is greater than the marginal benefit, the individual defects. Now the dilemma rears its ugly head: each individual faces the same choice; thus all defect and the common good is not produced at all. The individually rational strategy of weighing costs against benefits results in an inferior outcome and no common good is produced.

However, the logic behind the decision to cooperate or not changes when the interaction is ongoing since future expected utility gains will join present ones in influencing the rational individual’s decision. In particular, individual expectations concerning the future evolution of the game can play a significant role in each member’s decisions. The importance given the future depends on how long the individuals expect the interaction to last. If they expect the game to end soon, then, rationally, future expected returns should be discounted heavily with respect to known immediate returns. On the other hand, if the interaction is likely to continue for a long time, then members may be wise to discount the future only slightly and make choices that maximize their returns on the long run. Notice that making present choices that depend on the future is rational only if, and to the extent that, a member believes its choices influence the decisions others make.

One may then ask the following questions about situations of this kind: if agents make decisions on whether or not to cooperate on the basis of imperfect information about group activity, and incorporate expectations on how their decision will affect other agents, then how will the evolution of cooperation proceed? In particular, which behaviors are specific to the type of expectations and which are more general?

## *The economics of free riding*

In our mathematical treatment of the collective action problem we state the benefits and costs to the individual associated with the two actions of cooperation and defection, i.e., contributing or not to the social good. The problem thus posed, referred to in the literature as the *n*-person prisoner's dilemma [12, 8, 13], is presented below. In Section 3 we will then show how beliefs and expectations about other individuals' actions in the future can influence a member's perceptions of which action, cooperation or defection, will benefit him most in the long run. We will also discuss the different classes of expectations and convolve expectations into individual utility. Using these preferences functions, in Section 4 we will apply the stability function formalism [14] to provide an understanding of the dynamics of cooperation.

In our model of social dilemmas, each individual can either contribute (cooperate) to the production of the good, or not (defect). While no individual can directly observe the effort of another, each member observes instead the collective output and can deduce overall group participation using knowledge of individual and group production functions. We also introduce an amount of uncertainty into the relation between members' efforts and group performance. There are many possible causes for this uncertainty [13]; for example, a member may try but fail to contribute due to unforeseen obstacles. Alternatively, another type of uncertainty might arise due to individuals with bounded rationality occasionally making suboptimal decisions [15, 16]. In any case, we treat here only idiosyncratic disturbances or errors, whose occurrences are purely uncorrelated.

Consequently, we assume that a group member intending to participate does so successfully with probability $p$ and fails with probability $1-p$, with an effect equivalent to a defection. Similarly, an attempt to defect results in zero contribution with probability $q$, but results in unintentional cooperation with probability $1-q$. Then, as all attempts are assumed to be uncorrelated, the number of successfully cooperating members, $\widehat{n}_c$, is a mixture of two binomial random variables with mean $< \widehat{n}_c >= pn_c + (1-q)(n-n_c)$, where $n_c$ is the number of members attempting to cooperate within a group of size $n$.

Let $k_i$ denote whether member $i$ intends to cooperate ($k_i = 1$) or defect ($k_i = 0$), and let $k'_i$ denote whether member $i$ is cooperating or defecting in effect. The number of members cooperating is $\widehat{n}_c = \sum k'_j$. The limit $p$ and $q$ equal to 1 corresponds to an error-free world of complete information, while $p$ and $q$ equal to 0.5 reflect the case where the effect of an action is completely divorced from intent. Whenever $p$ and $q$ deviate from 1, the perceived level of cooperation will differ from the actual attempted amount.

In a simple, but general limit, collective benefits increase linearly in the contributions of the members, at a rate $b$ per cooperating member. Each contributing individual bears a personal cost, $c$. Then the utility at time $t$ for member $i$ is

$$U_i(t) = \frac{b}{n}\widehat{n}_c(t) - ck_i. \quad (1)$$

Using its knowledge of the functional form of the utility function,[1] each individual can deduce the number of individuals effectively cooperating from the utility collected at time $t$ by inverting Eq. 1:

$$\widehat{n}_c(t) = \frac{n}{b}(U_i(t) + ck_i). \quad (2)$$

This estimation differs from the actual number of individuals intending to cooperate in a manner described by the mixture of two binomial distributions. We also define $\widehat{f}_c(t)$, to denote the fraction, $\widehat{n}_c(t)/n$, of individuals effectively cooperating at time $t$.

When all members contribute successfully, each receives net benefits $(bn/n) - c = b - c$, independent of the group size. The production of the collective good becomes a dilemma when

$$b > c > \frac{b}{n}. \quad (3)$$

Thus, although the good of all is maximized when everyone cooperates ($b - c > 0$), the dominant strategy in the one-shot game is to defect since additional gain of personal participation is less than the private cost ($b/n - c < 0$).

[1] For a justification of the form of the individual utility function in the context of either divisible goods or pure public goods see [13].

## 3. Expectations and Beliefs

How agents take into account the future is wrapped into their expectations. The barest notion of expectations comes from the economic concept of horizon length. The horizon length is how far an agent looks into the future, or how long the agent expects to continue interacting with the other agents in the group. The horizon length may be limited by an agent's lifetime, by the agent's projection of the group's lifetime, by bank interest rates, etc.

In our framework, agents believe that their present actions will affect those of others in the future. In particular, the agents expect that defection encourages defection and cooperation encourages cooperation[2], but to a degree that depends on the size of the group and the present level of production. How agents believe their degree of influence to vary is left unspecified within the broader framework and is instead instantiated within the various classes of expectations. In addition, the larger the group, the less significance agents accords their actions: the benefit produced by an agent is diluted by the size of the group when it is shared among all agents. Agents that free ride can expect the effect to be very noticeable in a small group, but less so in a larger group. This is similar to the reasoning students might use when deciding whether or not to attend a lecture they would prefer to skip. Among an audience of 500, one's absence would probably go unnoticed (but if all students in the class reason similarly...). On the other hand, in a small seminar of ten, one might fear the personal censure of the professor.

This framework of expectations leaves unspecified how individuals believe their actions will affect others in the future. The specification of these beliefs can be covered by five different classes, which are represented qualitatively in Fig. 1 and quantitatively by the expectation function, $E(\widehat{f_c})$, a measure of individuals' perception of their influence on others. Low values of $E(\widehat{f_c})$ indicate that individuals believe their influence to be small so that however they act will have little effect on the future of the group. High values, on the other hand, indicate that individuals believe their influence to be large. These expectation functions shown are intended to be *representative* of the five different classes and are not to be thought of as constrained to the exact shape shown.

In general, agent expectations will be some mixture of these sets of beliefs, often with some class dominating. The first set of beliefs pictured in Fig. 1(a) we call flat expectations, according to which agents believe that the effect of their actions is independent of the proportion of the group cooperating. The "contrarian" and "inverse" classes of expectations shown in (d) and (e) we consider unrealistic because agents with these expectations believe that cooperation will induce cooperation at high rates even when the level of cooperation in the group is very low.

[2] This assumption is justified in part by Quattrone and Tversky's experimental findings in the context of voting that many people view their own choices as being diagnostic of the choices of others, despite the lack of causal connections [17].

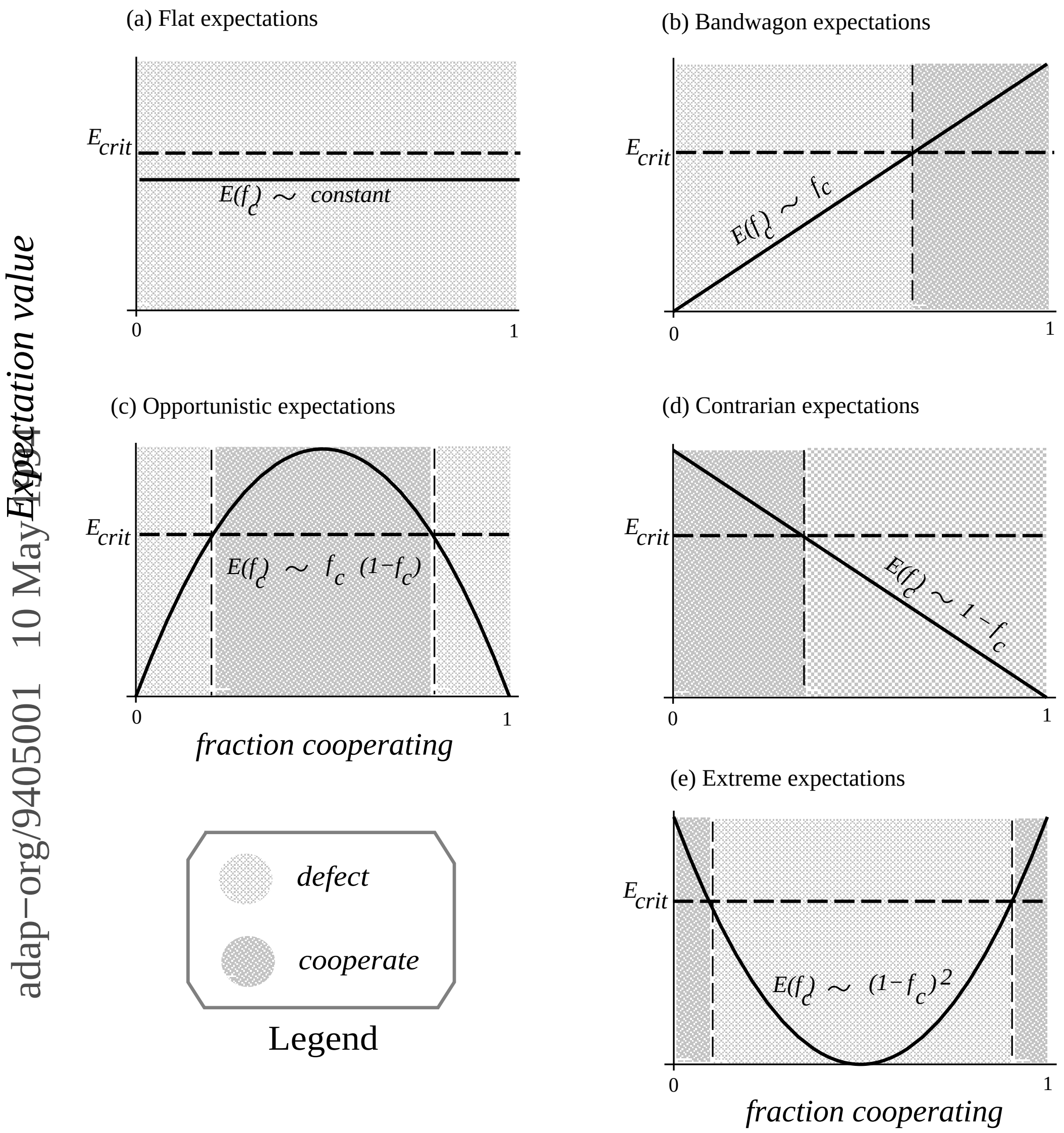

**Fig. 1.** The different classes of expectations that fit within our framework are (a) flat expectations, (b) bandwagon expectations, (c) opportunistic expectations, (d) inverse expectations, and (e) extreme expectations. These expectation functions shown are intended to be *representative* of the five different classes and are not to be thought of as constrained to the exact shape shown. The value of the expectation function, $E(\widehat{f}_c)$, indicates how strongly individuals believe that their actions will encourage similar behavior by the rest of the group. We emphasize bandwagon and opportunistic expectations. Agents with bandwagon expectations believe that the group will imitate their actions to the extent that its members are already behaving similarly. Agents with opportunistic expectations believe instead that when most of the group is cooperating, then the imitative surge in response to their actions will be small.

Fig. 1(b) shows a type of expectations, $E(\widehat{f_c}) \sim \widehat{f_c}$, we call "bandwagon" which assumes that agents believe that the group will imitate their actions to the extent that its members are already behaving similarly. Bandwagon agents expect that if they decide to free ride in a group of contributors, others will eventually choose to defect as well. The agents also believe that the rate at which the switchover occurs over time depends on the fraction of the group presently cooperating. The more agents already cooperating, the faster the expected transition to defection. Similarly, agents expect that if they start cooperating in a group of free riders, others will start cooperating over time. Once again the agents believe that the rate depends on the proportion of cooperators, which in this case is very low. The key assumption behind bandwagon expectations is that agents believe their actions influence the contributors more than the sluggards.

Consider the set of beliefs the agent expects of others in the context of recycling programs. Recycling has a strong public good component because its benefits are available to all regardless of participation. Not too long ago very few towns had such programs. Perhaps you would read in the paper that a small town in Oregon had started a recycling program. Big deal. But several years later, when you read that cities all over your state have jumped onto the recycling bandwagon, then suddenly the long-term benefits of recycling seem more visible: recycled products proliferate in the stores, companies turn green, etc. Alternatively, imagine some futuristic time when everyone recycles, in fact your town has been recycling for years, everything from cans to newspapers to plastic milk jugs. Then you hear that some places are cutting back their recycling efforts because of the expense and because they now believe that the programs don't do that much good after all. You think about all your wasted effort and imagine that the other towns still recycling are reaching the same conclusion. In view of this trend, your commitment to recycling may falter.

The "opportunistic" expectations, $E(\widehat{f_c}) \sim \widehat{f_c}(1 - \widehat{f_c})$, of Fig. 1(c) resemble bandwagon expectations when the proportion of cooperating members is small. However, when most of the group is cooperating, the agent believes that the imitative surge in response to his action will be small: the higher the fraction cooperating, the lower the expected response to an occasional defection or a marginal cooperation. Thus, the agent may be tempted to opportunistically defect, enjoying the fruits of cooperation without incurring the cost. Opportunistic expectations merges bandwagon expectations at the low end and contrarian expectations at the high end; when the amount of cooperation is small, opportunistic agents believe that their cooperation will encourage cooperation and their defection will encourage defection at a rate that depends the fraction cooperating, but that when the level of cooperation of high, the predicted rate depends on the fraction defecting instead.

Opportunistic expectations apply, for example, to situations in which agents believe that the common good can be produced by a subset of the group (even when that is not the case). Consider a common good such as public radio. The more voluntary contributions the radio station receives, the better the quality of the programming. An opportunistic listener who perceives that the station's received voluntary contributions are adequate will decide not to send his yearly contribution believing that his own action has little consequence.

*Expectations and utility*

In Section 2 we presented the utility function for an individual faced with a social dilemma, neglecting the effect of expectations. We now return to the utility calculation to include the influence of expectations on preferences. This can be done in a fairly general way using the expectation function, $E(\widehat{f_c})$. Individuals use their expectation function to extrapolate perceived levels of cooperation into the future. The framework underlying the different classes of expectations assumes that the members of the group expect the game to be of finite duration, parametrized by their horizon length, $H$.[3] Since finite horizons mean that a dollar today is worth more than a dollar tomorrow, agents discount future returns expected at a time $t'$ from the present at the rate $e^{-t'/H}$ with respect to immediate expected returns. Secondly, members expect that their choice of action, when reflected in the net benefits received by the others, will influence future levels of cooperation. Since, however, the decision of one individual affects another's return by only $\pm b/n$, we assume that members perceive their influence as decreasing with increasing group size.

The time scale of the dynamics is normalized using the parameter $\alpha$, which is the rate at which members of the group reexamine their choices. Individuals deduce the level of cooperation, $\widehat{f_c}$, from their received utility, as per Eq. 2. Since the amount of utility obtained may depend on past (instead of present) levels of cooperation, the deduced fraction cooperating may actually correspond to a previous state of the group. We use the parameter $\tau$ to represent this delay in information. The deduced fraction cooperating also differs from the past value because of uncertainty, as discussed earlier.

Below, we develop mathematically how individual expectations convolves with instantaneous utility to determine the condition for cooperation. Let $t$ represent the present time and let $\Delta\widehat{f_c}(t+t')$ denote the expected future difference (at time $t+t'$) between the fraction of agents cooperating and the fraction of those defecting. Fig. 2 shows how agents exptrapolate the observed aggregate behavior of the group into the future. Because of delays, the agents do not know the evolution of the dynamics in the gray area of the figure. Instead they use the delayed value of the fraction cooperating, $f_c(t-\tau)$, to extrapolate the group's behavior into the present and then into the future.

[3] The concept of a horizon is formally related to a discount $\delta$, which reflects the perceived probability that the game will continue through the next time step. The two are connected through the relation $\sum_{i=0}^{\infty} \delta^i \,() \to \int_0^{\infty} dt'\, e^{-t'/H}\,()$ which implies $H = 1/(1-\delta)$.

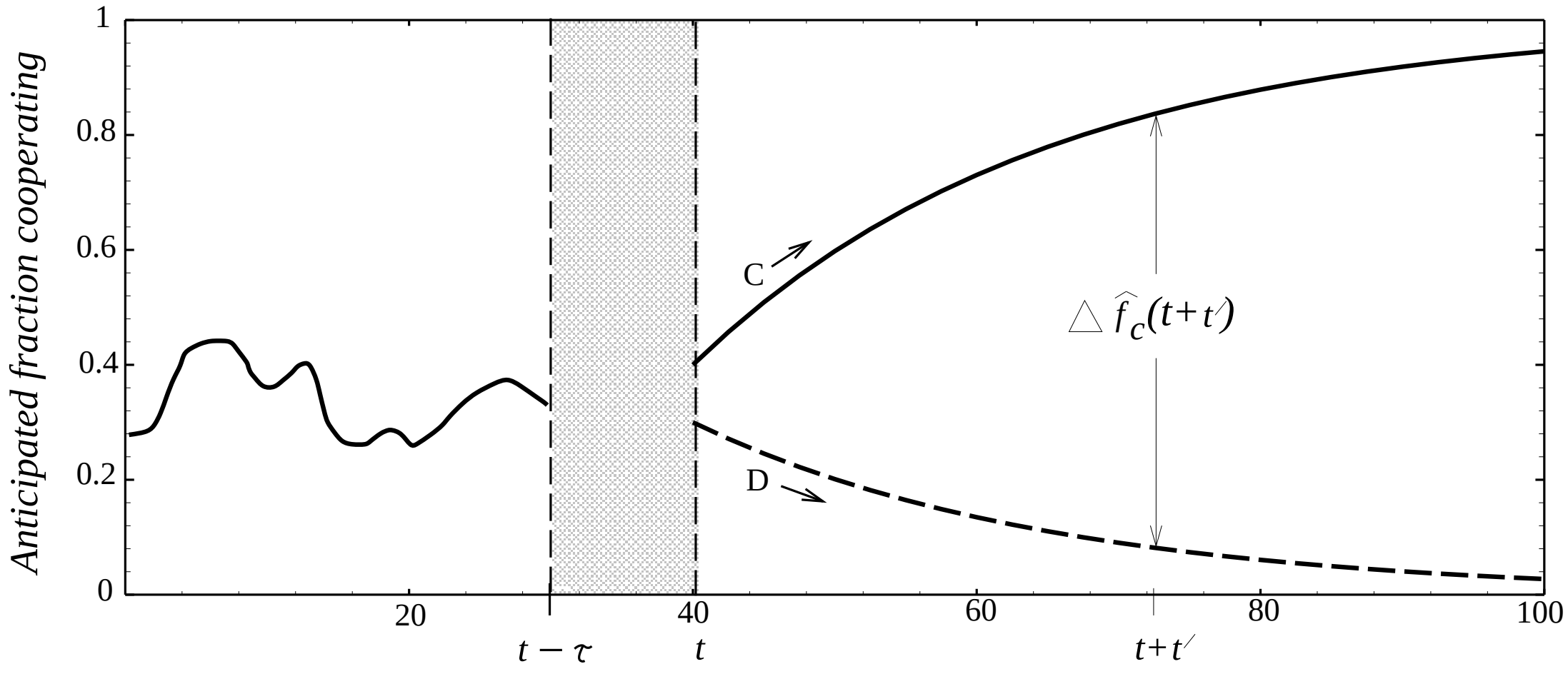

**Fig. 2.** Agents extrapolate the evolution of the group's behavior using delayed information of the fraction cooperating. The upper curve represents how a member expects the dynamics to evolve if he cooperates, the lower if he defects. The rate with which the two curves respectively rise and fall depends on the members' class of expectations functions. Agents with bandwagon expectations extrapolate out very flat curves when the fraction cooperating is small and much steeper curves when this fraction is large. Opportunistic agents, in contrast, extrapolate out very flat curves whenever the fraction cooperating becomes either small or large, but predict sharper rises and declines for intermediate levels of cooperation.

The difference between the two curves in Fig. 2 corresponds to the time-varying expected deviance $\Delta \widehat{f}_c(t+t')$. The upper curve in the figure represents a member's extrapolation of group behavior if he cooperates, the lower curve if he defects. The initial slope of the curves is proportional to the individual's expectation function. Thus, when the expectation function has a small value, the expected deviation grows very slowly — representing the individual's belief that one's actions count for little. On the other hand, when the expectation function yields a large value, the curves respectively rise and fall rapidly — indicating the belief that one's action influence others strongly. The rate at which the extrapolated curves rise and fall also depends on the size of the group and the reevaluation rate: the first because of the assumption that influence declines with group size and the second because $\alpha$ sets the time scale.

Because a member's choice causes an instantaneous difference at $t' = 0$ of $\Delta \widehat{f}_c(t, t' = 0) = 1/n$, the extrapolated deviation at time $t + t'$ becomes

$$\Delta \widehat{f}_c\big(t+t'\big) = 1 - (1 - 1/n)\exp\left(-\frac{\alpha E\Big(\widehat{f}_c(t-\tau)\Big)}{n}t'\right). \tag{4}$$

In summary, members reevaluate their decision whether or not to contribute to the production of the good at an average rate $\alpha$, deducing the value $\widehat{f}_c(t-\tau)$ and following some set expectations about the future, represented by the expectation function, $E(\widehat{f}_c)$.

From the members' prediction of how they expect $f_c$ to evolve in relation to their choice and discounting the future appropriately, they then make the decision on whether to cooperate or defect by estimating their expected utility over time.

Putting it all together, individuals perceive the advantage of cooperating over defecting at time $t$ to be the net benefit

$$\begin{aligned} \Delta \widehat{B}_i(t) &= \int_0^\infty dt' e^{-t'/H} \left\{ b\,\Delta \widehat{f}_c(t+t') - c \right\} \\ &= H(b-c) - \frac{Hb(n-1)}{n + H\alpha E\left(\widehat{f}_c(t-\tau)\right)}. \end{aligned} \tag{5}$$

An individual cooperates when $\Delta \widehat{B}_i(t) > 0$, defects when $\Delta \widehat{B}_i(t) < 0$, and chooses at random between defection and cooperation when $\Delta \widehat{B}_i(t) = 0$. The decision is based on the fraction of the group perceived as cooperating at a time $\tau$ in the past, $\widehat{f}_c(t-\tau)$. These criteria reduce to the following condition for cooperation at time $t$:

$$E_{crit} \equiv \frac{1}{H\alpha}\left(\frac{nc-b}{b-c}\right) < E\left(\widehat{f}_c(t-\tau)\right). \tag{6}$$

The condition for cooperation derived above depends explicitly on the class of expectations and can be visualized using the Fig. 1. In the absence of uncertainty an agent with bandwagon expectations cooperates in the dark gray region where

$$E_{crit} < m\widehat{f}_c(t-\tau), \tag{7}$$

while an opportunistic agent cooperates in the dark gray region of (c) where

$$E_{crit} < m\widehat{f}_c(t-\tau)\left(2f_c^{max} - \widehat{f}_c(t-\tau)\right). \tag{8}$$

The parameter $m$ indicates the steepness of the expectation functions and $f_c^{max}$ indicates the level of cooperation at which opportunistic expectations reach a maximum. Fig. 1(b) and (c) should help make these conditions for cooperation more clear: in the cross-hatched regions where $E(f_c) > E_{crit}$ agents will prefer cooperation (neglecting imperfect information), while in the lightly-shaded regions where $E(f_c) < E_{crit}$ agents will prefer defection.

Since $\widehat{f}_c(t-\tau)$ is a mixture of two binomially distributed variables, Eq. 6 provides a full prescription of the stochastic evolution for the interaction, depending on the relevant class of expectations. In particular, members cooperate with probability $P_c(f_c(t-\tau))$ that they perceive cooperation as maximizing their expected future accumulated utility, given the actual attempted level of cooperation $f_c(t-\tau)$.

## 4. Theory

As in [1] we borrow methods from statistical thermodynamics [18] in order to study the evolution of social cooperation. This field attempts to derive the macroscopic properties of matter (such as liquid versus solid, metal or insulator) from knowledge of the underlying interactions among the constituent atoms and molecules. In the context of social dilemmas, we adapt this methodology to study the aggregate behavior of a group composed of intentional individuals confronted with social choices.

*Dynamics*

A differential equation describing the stochastic discrete interaction specified in the previous section can be derived using the mean-field approximation. This entails assuming that: (1) the size, *n*, of the group is large; and (2) the average value of a function of some variable is well approximated by the value of the function at the average of that variable. The equations developed below will allow us to determine the equilibrium points and their stability characteristics.

We specialize to the symmetric case $p = q$ in which an individual is equally likely to be perceived as defecting when he intended to cooperate as cooperating when he intended to defect. By the Central Limit Theorem, for large *n*, the random variable $\widehat{f_c}$ tends to a Gaussian distribution with mean $\langle \widehat{f_c} \rangle = pf_c + (1-p)(1-f_c)$. Using the distribution of $\widehat{f_c}$, we now want to find the mean probability, $\langle \rho_c(\widehat{f_c}) \rangle$, that $E(\widehat{f_c}) > E_{crit}$. Within the mean-field approximation, $\langle \rho_c(\widehat{f_c}) \rangle$ is given by the probability $\rho_c(f_c)$ that $E(\langle \widehat{f_c} \rangle) > E_{crit}$, (*i.e.*, $\langle \rho_c(\widehat{f_c}) \rangle = \rho_c[E(\langle \widehat{f_c} \rangle)]$ ). Thus, the mean probability that $E(\widehat{f_c}) > E_{crit}$ becomes

$$\rho_c(f_c) = \frac{1}{2}\left\{1 + \text{erf}\left[\frac{E(\langle \widehat{f_c} \rangle) - E_{crit}}{\sqrt{2}\sigma}\right]\right\}, \tag{9}$$

where $\text{erf}\left(x/\sqrt{2}\sigma\right)$ represents the error function associated with the normal curve. The uncertainty parameter $\sigma$ captures imperfect information and the spread of diversity of beliefs within the group [2]. The evolution of the number of agents cooperating in time is then described by the dynamical equation [19]

$$\frac{df_c}{dt} = -\alpha[f_c(t) - \rho_c(f_c(t-\tau))], \tag{10}$$

where $\alpha$ is the reevaluation rate and $\tau$ is the delay parameter, as defined earlier.

Fig. 3 pictorializes this differential equation for the bandwagon expectations in (a) and for opportunistic ones in (b). The figures superimpose $y = \rho_c(f_c)$ and $y = f_c$.

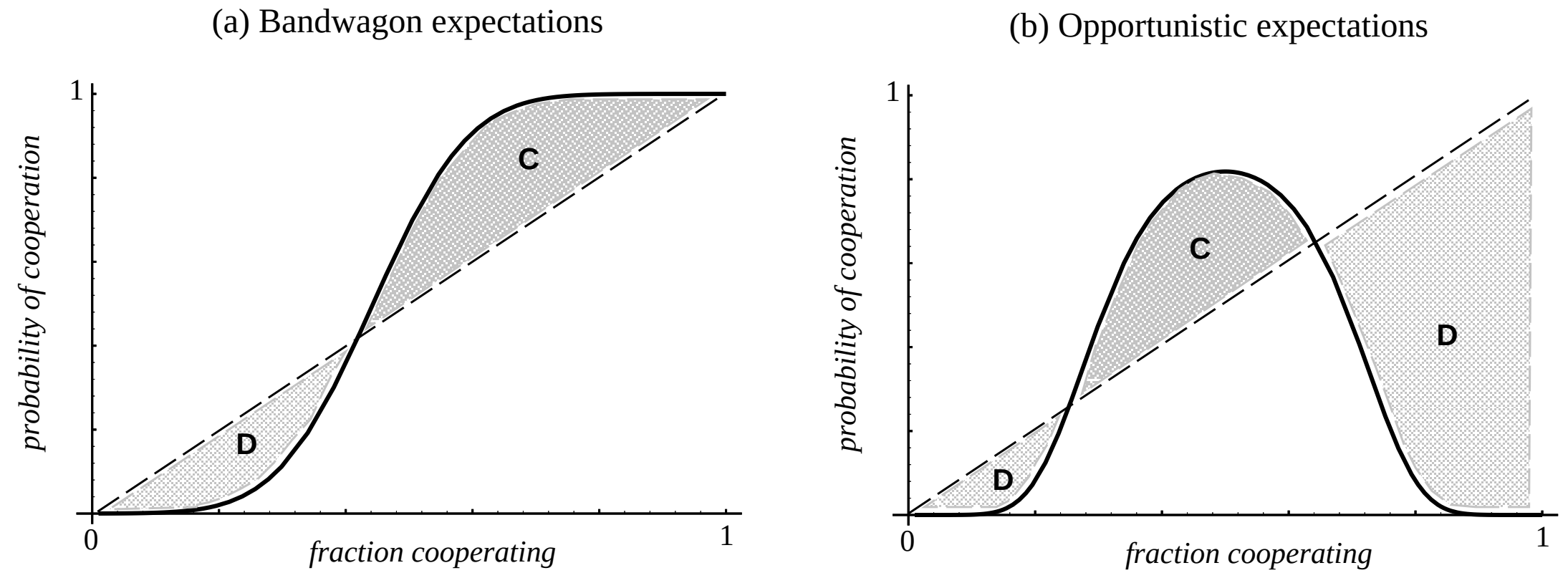

**Fig. 3.** The fixed points of the differential equation describing the dynamics of group cooperation are given by the intersection of $y = \rho_c(f_c)$ and $y = f_c$. In region C where $\rho_c(f_c) > f_c$, the dynamics evolves towards more cooperation, while in region D where $f_c > \rho_c(f_c)$, the dynamics evolves towards more defection. Contrasting (a) and (b) hints at the difference in the dynamical behavior for bandwagon vs. opportunistic expectations. For opportunistic expectations, the more cooperative fixed point occurs where $\rho_c(f_c)$ intersects $f_c$ with a negative slope. When the slope at this point of intersection is steep enough, the fixed point becomes unstable causing oscillations and chaos, behavior that is not observed for bandwagon expectations.

When the difference $f_c - \rho_c(f_c)$ is positive, group members prefer defection (region D), when it is negative they prefer cooperation (region C). The fixed points of the differential equation are given by the points of intersection of the two curves; the second of the three fixed points is unstable since a small cooperative perturbation takes the system into the more cooperative equilibrium while a small perturbation in the other direction takes the system into the fixed point of overall defection.

As the size of the group increases, the curve $y = \rho_c(f_c)$ shifts so that it intersects $y = f_c$ only near $f_c = 0$. By solving

$$\rho(f_0) = f_0, \tag{11}$$

we can obtain the critical sizes beyond which cooperation can no longer be sustained for the different classes of expectations. In the case of perfect certainty ($p = q = 1$), these critical sizes can be expressed in simple analytical form. Consider first an instantiation of bandwagon expectations:

$$E(\widehat{f_c}) = \widehat{f_c}. \tag{12}$$

In this case, the critical group size beyond which cooperation is no longer a fixed point occurs when $E_{crit} > 1$, i.e., for groups of size greater than

$$n^* = H\alpha\left(\frac{b}{c} - 1\right) + \frac{b}{c}. \tag{13}$$

This can be seen both pictorially in Fig. 1(b) and from Eq. 6: with perfect information, no individual will cooperate when $E_{crit}$ is greater than the expectation function $E(\widehat{f}_c)$ for all $f_c$. Similarly, if we choose the opportunistic expectations

$$E\left(\widehat{f}_c\right) = 4\widehat{f}_c\left(1 - \widehat{f}_c\right) \tag{14}$$

the solution to Eq. 10 yields only one non-cooperative fixed point for $n > n^*$.

Cooperation is the only possible global outcome if the group size falls below a second critical size $n_{min}$. For perfect information,

$$n_{min} = \frac{b}{2c} + \frac{1}{2c}\sqrt{b^2 + 4H\alpha c(b - c)} \tag{15}$$

for bandwagon agents and

$$n_{min} = \frac{b}{2c} + \frac{1}{2c}\sqrt{b^2 + 16H\alpha c(b - c)} \tag{16}$$

(approximately) for opportunistic agents. Notice that in either case these two critical sizes are not equal; in other words, there is a range of sizes between $n_{min}$ and $n^*$ for which both cooperation and defection are possible fixed points.

An estimate of the possible critical group sizes can be obtained if one assumes, for example, a horizon length $H = 50$ (which corresponds to a termination probability $\delta = 0.98$), the reevaluation rate $\alpha = 1$, the benefit for cooperation $b = 2.5$ and the cost of cooperation $c = 1$. In this case one obtains $n^* = 77$, $n_{min} = 10$ in the case of bandwagon expectations and $n_{min} = 18$ for opportunistic agents. Observe that an increase in the horizon length would lead to corresponding increases in the critical sizes.

Finally, linear stability analysis of the differential equation (Eq. 10) shows that for bandwagon expectations the stability of the equilibrium points is independent of the value of the delay $\tau$ and of the reevaluation rate $\alpha$. Thus, the asymptotic behavior of the group interaction does not depend on the delay; this observation is corroborated for the discrete model by numerical computer simulations such as those presented in Section 5. Moreover, the equilibrium points belong to one of two types: stable fixed point attractors or unstable fixed point repellors. Due to the linearity of the condition for cooperation for bandwagon expectations (Eq. 6), the dynamical portrait of the continuous model contains no limit cycles or chaotic attractors.

However, for opportunistic expectations the more cooperative equilibrium point becomes unstable for large delays, although the fixed point near $f_c = 0$ remains stable. Integrating the differential equation above reveals a panoply of dynamical behavior from oscillations to chaos. For small delays, the dynamics of opportunistic expectations is very similar to that of bandwagon expectations: two stable fixed points separated by a barrier. We now examine how the stochastic fluctuations arising from uncertain information affects the dynamics of social dilemmas.

### *Fluctuations*

Eq. 10 determines the average properties of a collection of agents having to choose between cooperation and defection. The asymptotic behavior generated by the dynamics provides the fixed points of the system and an indication of their stability; it does not address the question of how fluctuations away from equilibrium state evolve in the presence of uncertainty. These fluctuations are important for two reasons: (1) the time necessary for the system to relax back to equilibrium after small changes in the number of agents cooperating or defecting might be long compared to the time-scale of the collective task to be performed or the measuring time of an outside observer; (2) large enough fluctuations in the number of defecting or collaborating agents could shift the state of the system from cooperating to defecting and vice-versa. If that is the case, it becomes important to know how probable these large fluctuations are, and how they evolve in time.

In what follows we will use a formalism developed in [14] that is well suited for studying fluctuations away from the equilibrium behavior of the system. This formalism relies on the existence of an optimality function, $\Omega$, that can be constructed from knowledge of the utility function. The $\Omega$ function has the important property that its local minima are the equilibria of the system as well as the most probable configurations of the system. Depending on the complexity of the $\Omega$ function, there may be several equilibria, with the overall global minimum being the optimal state of the system.

Specifically, the equilibrium probability distribution $P_e(f_c)$ is given by

$$P_e = C \exp\left[-n\Omega(f_c)\right], \tag{17}$$

where the optimality function $\Omega$ for our model of ongoing collective action is given by

$$\Omega(f_c) = \int_0^{f_c} df' \left[f' - \rho_c(f')\right] \tag{18}$$

in terms of the mean probability $\rho_c(f_c)$ that cooperation is preferred. Thus, the optimal configuration corresponds to the value of $f_c$ at which $\Omega$ reaches its global minimum.

Within this formalism it is easy to study the dynamics of fluctuations away from the minima in the absence of delays. First, consider the case where there is a single equilibrium (which can be either cooperative or defecting). Fluctuations away from this state relax back exponentially fast to the equilibrium point, with a characteristic time of the order of $1/\alpha$, which is the average reevaluation time for the individuals. Alternatively, there may be multiple equilibria, with the optimal state of the system given by the global minimum of the $\Omega$ function. A situation in which the system has two equilibria is schematically illustrated in Fig. 4 for both bandwagon and opportunistic expectations. In (a) an $\Omega$ function characteristic of bandwagon expectations is shown whose minima are at either extreme and in (b) an $\Omega$ function characteristic of opportunistic expectations is shown with one minima at $f_c \approx 0$ and a second minima occurring at an intermediate value of $f_c$. The quadratic form of this second minima accounts for the discrepancy in the dynamics of opportunistic expectations versus bandwagon expectations.

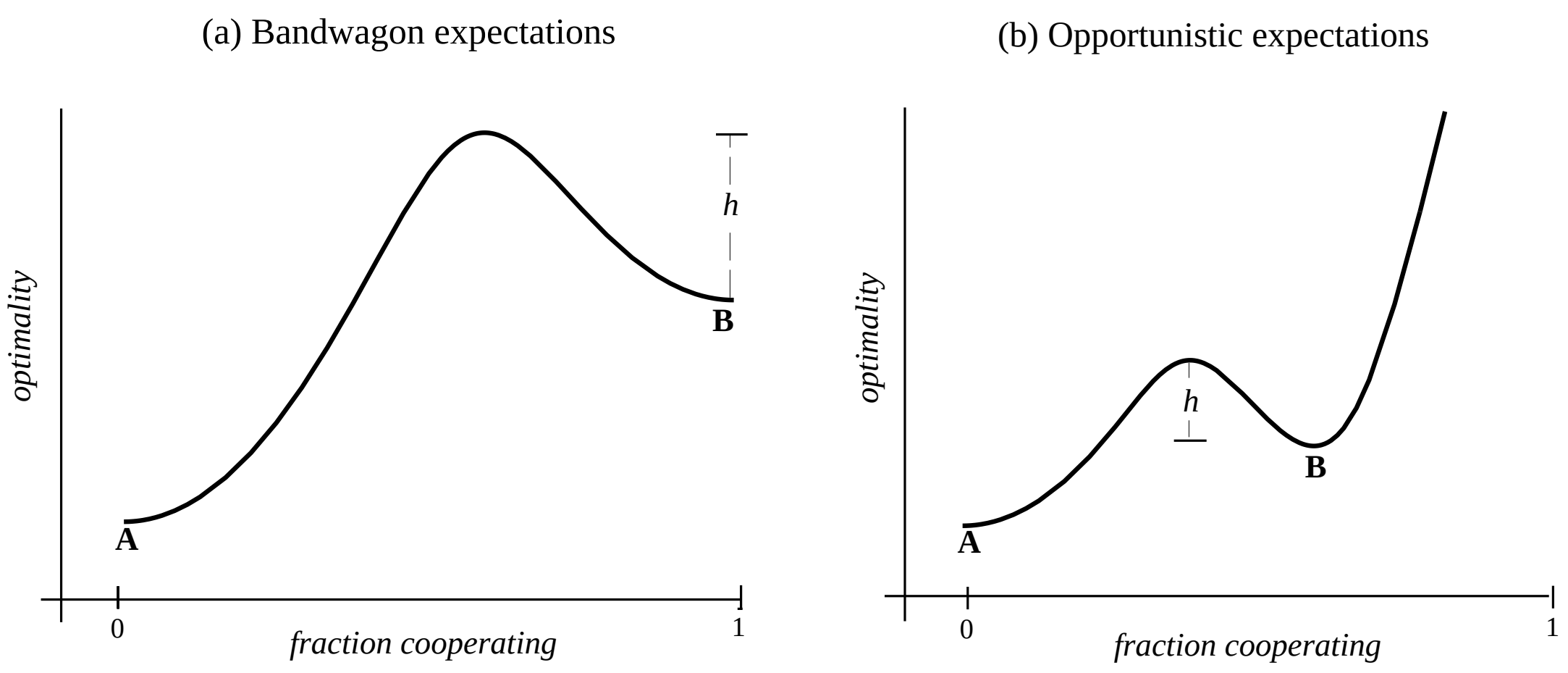

**Fig. 4.** The optimality function $\Omega$ vs. $f_c$, the fraction of agents cooperation, for bandwagon expectations in (a) and opportunistic expectations in (b). The global minimum is at *A*, local minimum at *B*, and *h* is the height of the barrier separating state *B* from *A*.

If the system is initially in an equilibrium which corresponds to the global minimum (e.g., state *A*), fluctuations away from this state will relax back exponentially fast to that state. But if the system is initially trapped in a metastable state (state *B*), the dynamics away from this state is both more complicated and interesting. As was shown in [14], within a short time scale, fluctuations away from a local minimum relax back to it, but within a long time scale, a giant fluctuation can take place in which a large fraction of the agents switches strategies, pushing the system over the barrier maximum. Once the critical mass required for a giant fluctuation accumulates, the remaining agents rapidly switch to the new strategy and the system slides into the global equilibrium.

The time scales over which the nucleation of a giant fluctuation occurs is exponential in the number of agents. However, when such transitions take place, they do so very rapidly — the total time it takes for all agents to cross over is logarithmic in the number of agents. Since the logarithm of a large number is very small when compared to its exponential, the theory predicts that nothing much happens for a long time, but when the transition occurs, it does so very rapidly.

The process of escaping from the metastable state depends on the amount of imperfect knowledge that individuals have about the state of the system, in other words, on what individuals think the other agents are doing. In the absence of imperfect knowledge the system would always stay in the local minimum downhill from the initial conditions, since small excursions away from it by a few agents would reduce their utility. Only in the case of imperfect knowledge, which causes occasional large errors in the individual's estimation of the actual number cooperating, can a critical mass of individuals change their behavior.

Determining the time that it takes for the group to crossover to the global minimum is a calculation analogous to particle decay in a bistable potential and has been performed many times [18, 20]. The time, $t$, that it takes for a group of size $n$ to cross over from a metastable Nash equilibrium to the optimal one is given by

$$t = \text{constant } e^{nh/\sigma}, \tag{19}$$

with $h$ the height of the barrier as shown in Fig. 4 and $\sigma$ a measure of the imperfectness of information. We should point out, however, that in our model the barrier height itself also depends on $n$, $H$, and $p$, making simple analytical estimates of the crossover time considerably more difficult.

## 5. Computer Simulations

The theory presented in the previous section has a number of limitations. The mean-field dynamics provides an approximation to the model in the extreme limit of infinite group size, as does the $\Omega$ function formalism. In addition, the theory can predict whether or not the equilibrium points are stable, but not the characteristics of the dynamics when they are unstable. Finally, although the $\Omega$ function formalism predicts sudden transitions away from the metastable state into the global equilibrium and shows that the average time to transition is exponential in the group size, in order to calibrate the transition time, we must run computer experiments. While the theory may tell us that cooperation is the overall equilibrium point for a group of size 10 with horizon length 12, it does not indicate on average how long it will take for the system to break out of an initially defecting state (only that this time is exponential in the size of the group).

With this in mind, we ran a number of event-driven Monte Carlo simulations that we used to both test the analytical predictions of the theory and to calibrate the time constants with which cooperation and defection appear in a system of given size. These simulations run asynchronously: agents wake up at random intervals at an average rate $\alpha$ and reevalute their decision to cooperate or defect. They deduce the fraction cooperating from their accrued utility. Because of uncertainty, their deduction may differ from the true value cooperating. Higher levels of uncertainty result in large deviations from the true value. The agents then decide whether or not to cooperate according to the choice criterion given by Eq. 6. This choice criterion varies, of course, depending on the appropriate class of expectations. We present below a few representative results among the wide range of group dynamics and statistics collected for many different group sizes, horizon lengths and uncertainty levels

### *Fixed points and fluctuations*

The dynamics of bandwagon and opportunistic expectations are very similar when the delay in information vanishes (i.e., the limit $\tau \to 0$). In this case, as shown in the previous sections, our model of the collective action problem yields dynamics with two fixed points for a large range of group sizes, one an optimal equilibrium, the other a metastable equilibrium. If the group initially finds itself at or near the metastable equilibrium, it may be trapped there for very long times, making it effectively a stable state when the time scale of the interaction is shorter than the crossover time to the optimal configuration. This crossover time, $t$, given in Eq. 19, is exponential in the size $n$ of the group, in the height of the barrier, $h$, and in the inverse uncertainty, $1/\sigma$. Note that the barrier height is also a function of the uncertainty, $\sigma$, and indirectly of $n$ and the horizon length $H$, so the functional dependencies for the crossover time are more complex than appears at first sight.

Nevertheless, as verified through simulations of groups of individuals engaged in collective action problems, crossover times are exponentially long. As a result, a large group which has an initial tendency to cooperate may remain in a cooperative state for a very long time even though the optimal state is defection. Conversely, small groups whose initial tendency is to defect can persist in non-optimal defection, until a large fluctuation finally takes the group over to cooperation.

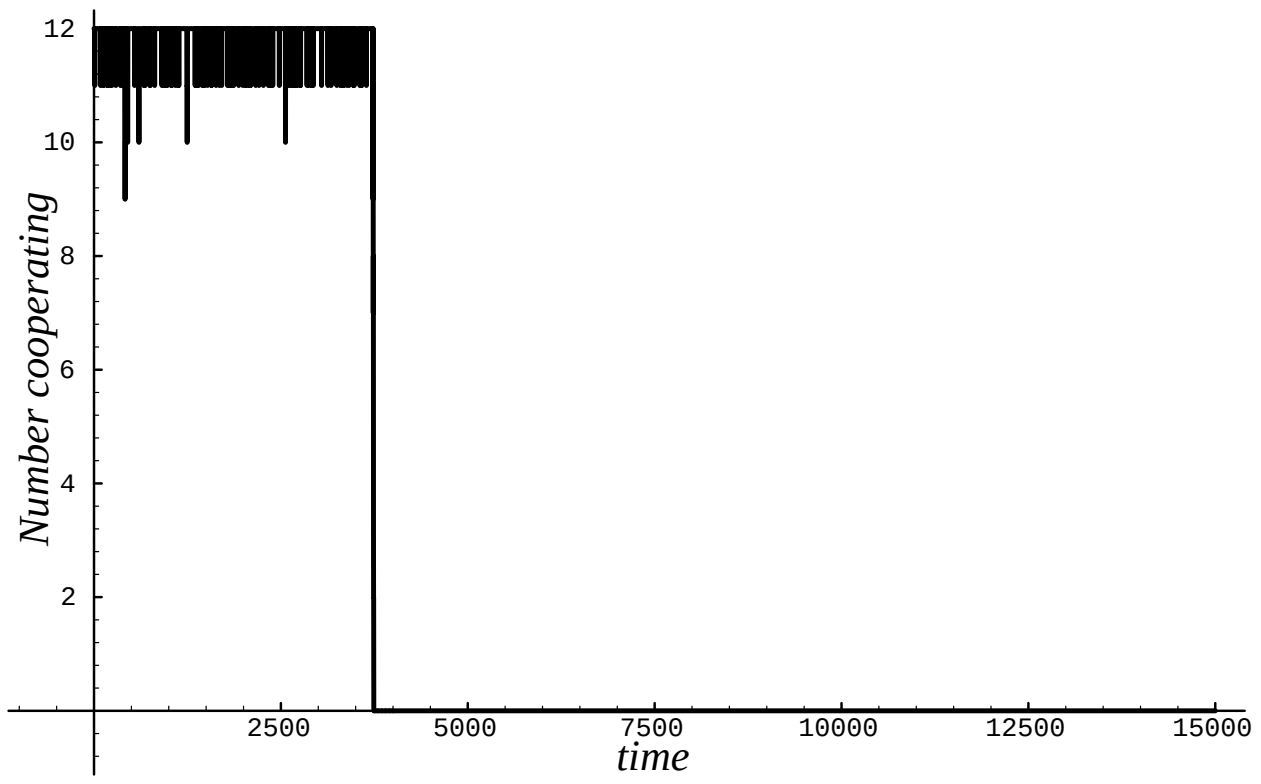

**Fig. 5.** *Outbreak of defection.* At $t = 0$, two cooperating groups of size $n = 6$ merge to form a larger, cooperating group of size $n = 12$. The agents follow bandwagon expectations and have horizon length $H = 9.5$ throughout, with $p = 0.93$, $b = 2.5$, $c = 1$, $\alpha = 1$ and $\tau = 1$. For these parameters, cooperation is the optimal state for a group of size $n = 6$, but for the combined group of size $n = 12$, cooperation is metastable. Indeed, as the figure shows, metastable cooperation persists for almost 4,000 time steps in this example. Uncertainty ($p$ less than one) ensures that eventually a large fluctuation in the perceived number of agents cooperating eventually takes the group over into a state of mutual defection, which is optimal.

As a concrete example, consider two small cooperating groups of size $n = 6$, with horizon length $H = 9.5$ and bandwagon expectations, for which the optimal state is cooperation. At $t = 0$, the groups merge to form a larger, cooperating group of size $n = 12$. For the larger group, cooperation is now a metastable state: no one individual will finds it beneficial to defect and the metastable cooperative state can be maintained for very long times, especially if $p$ is close to 1. As shown in Fig. 5, in one case mutual cooperation lasts for about 4,000 time steps, until a sudden transition (of duration proportional to the logarithm of the size of the group) to mutual defection occurs, from which the system will almost never recover (the time scale of recovery is many orders of magnitude larger than the crossover time). In this example, $p = 0.93$. If the amount of error increases so that $p$ now equals 0.91 (thus reducing the height of barrier between cooperation and defection by 21%), the crossover to defection occurs on the order of hundreds of time steps, instead of thousands.

Groups of agents with opportunistic instead of bandwagon expectations, behave qualitatively similarly in the absence of information delays. For a range of group sizes, there are again two equilibrium points, however, as predicted by the $\Omega$ function (e.g. Fig. 4(b)), the more cooperative equilibrium is actually a mixture of cooperation and defection. Thus, cooperation and defection can coexist with some agents free riding while other agents contribute. The dynamics, however, is ergodic as long as the agents have identical preferences and beliefs: all individuals "take turns" cooperating and defecting. Adding a bit of diversity, on the other hand, is enough to disrupt this parity: with diversity, at the mixed equilibrium point, some individuals will always tend to free ride while others contribute.

A second difference between opportunistic and bandwagon expectations at zero delays is that the typical size of fluctuations at the more cooperative equilibrium are larger for opportunistic groups. The quadratic curvature of the $\Omega$ function (Fig. 4(b)) at the second equilibrium explains the increase in fluctuations. For systems with bandwagon expectations, fluctuations from the cooperative equilibrium can take the system away from equilibrium in only one direction. (Although with high uncertainty the equilibrium points move away from the extremes; however, high uncertainty also washes away the difference between opportunistic and bandwagon expectations in more general ways.)

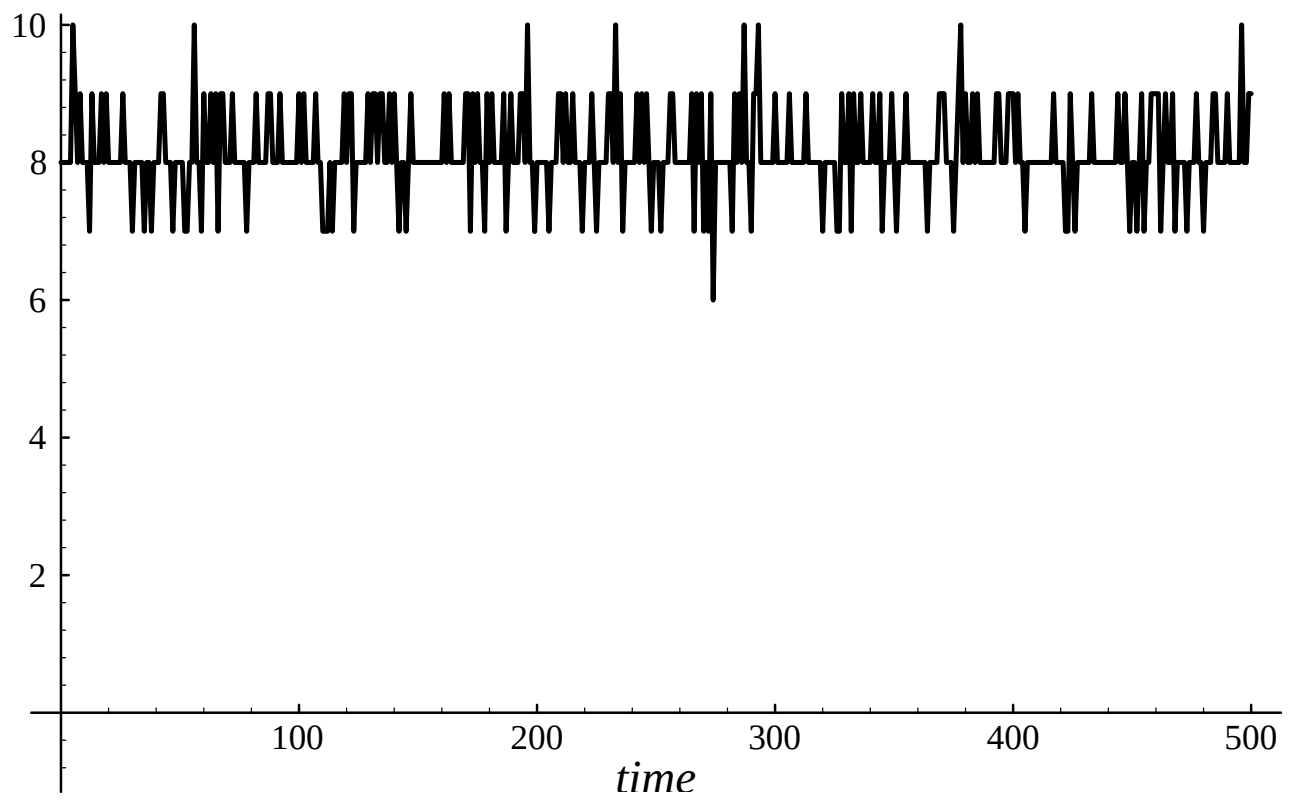

**Fig. 6.** Fluctuations about the mixed cooperative/defecting fixed point for an opportunistic group of size $n = 10$, horizon length $H = 10$, with $p = 0.95$. The number cooperating fluctuates around $n = 8$ and the typical size of fluctuations is $\pm 1$ with an occasional fluctuation of size two.

For example, the following opportunistic group has a long-term equilibrium which is a mixture of cooperation and defection: group size $n = 10$, horizon length $H = 10$, with $p = 0.95$. In this case, the number cooperating fluctuates around $n = 8$ and the typical size of fluctuations is $\pm 1$ with an occasional fluctuation of size two.

We also collected statistics to verify the exponential dependence of the mean crossover time on the probability $p$ that an agent's action is misperceived. For the data in Fig. 7, the crossover time from metastable cooperation to overall defection was averaged over 600 simulations at each value of the uncertainty parameter *p*. Fitting an exponential to the data indicates that the mean crossover time goes as $t \propto e^{f(p)}$, where $f(p)$ is quadratic in *p*. However, the more relevant relationship is between the mean crossover time and the height of the barrier of the $\Omega$ function. The mean crossover time *t* appears to be given by $t = \texttt{constant} \exp\left(h(n,\sigma)\right)$, where the barrier height depends on the group size, *n*, and the uncertainty, $\sigma$. For the example in Fig. 7, the barrier height depends quadratically on the parameter, *p*. The dependence of barrier height on group size is less straightforward, but apart from some outliers, the barrier height also increases quadratically in group size.

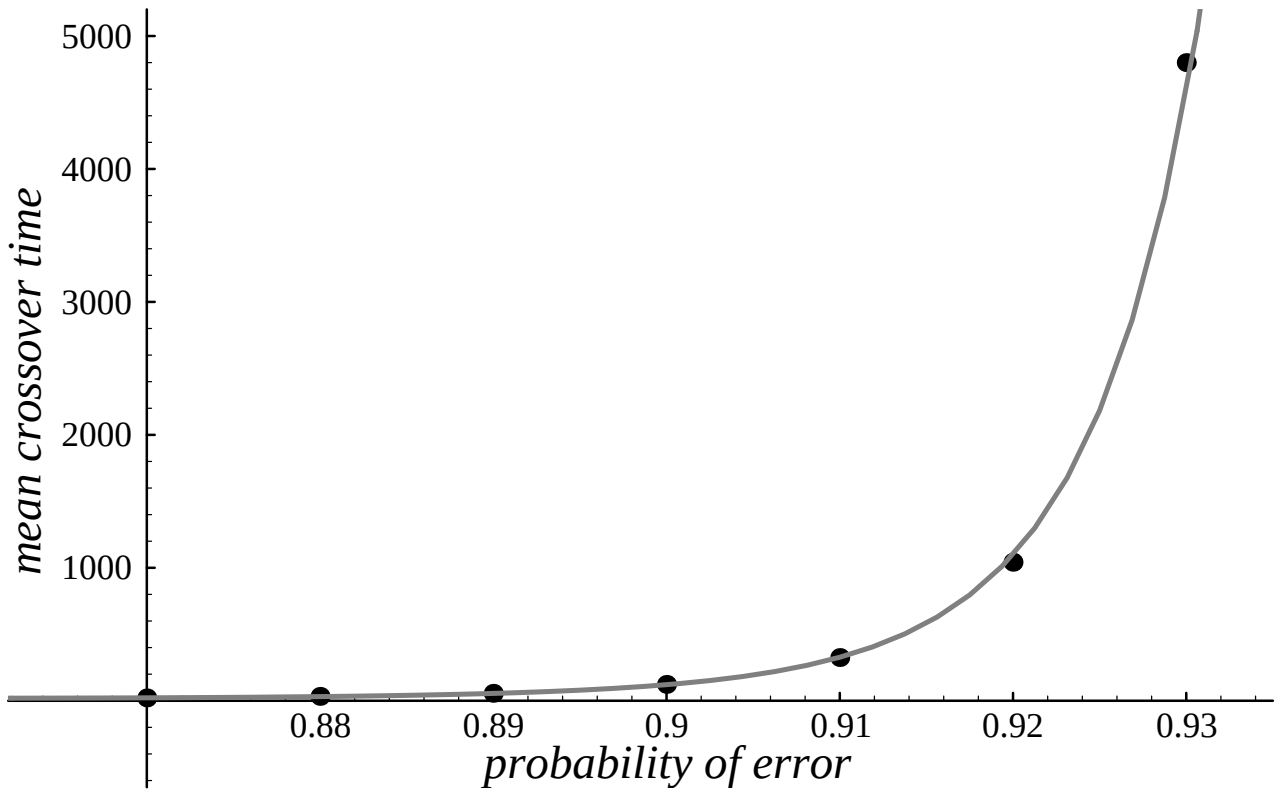

**Fig. 7.** *Exponential increase of crossover time* as a function of the probability *p* that an agent's action is misperceived. The mean crossover time was obtained by averaging over 600 simulations at each value of the uncertainty parameter *p*. In the all the runs, $n = 12$, $H = 9.5$, $b = 2.5$, $c = 1$, $\alpha = 1$ and $\tau = 1$, with bandwagon expectations. The gray curve is an exponential fit to the data.

## *Opportunistic oscillations*

When there are significant delays in information (in units of $1/\alpha$), the dynamical behavior of opportunistic agents becomes very different from that of agents with bandwagon expectations, in particular for small amounts of uncertainty. For large delays and low uncertainty, the equilibrium points of the system become unstable. However, for any given group of agents with opportunistic expectations, the dynamics can be stabilized regardless of the delay by simply increasing the amount of uncertainty (or, alternatively, adding diversity). This claim can be verified by performing a stability analysis about the fixed points of Eq. 10 and has been confirmed through computer experiments.

For long delays and low uncertainty, we observe a number of different behaviors. For example, when the agents' horizon length is long enough that the mixed equilibrium becomes more likely than the defecting one with a high barrier between the two, we observe large oscillations at low levels of uncertainty. Fig. 8 shows the time series of a computer experiment run in this regime. The size of the group is $n = 10$, the horizon length is $H = 10$, the delay is $\tau = 1$ and $p = 0.99$. Because of the small number of agents in the group, the stochastic dynamics is not purely oscillatory as predicted by the differential equation of Eq. 10, although the approximation improves for larger groups.

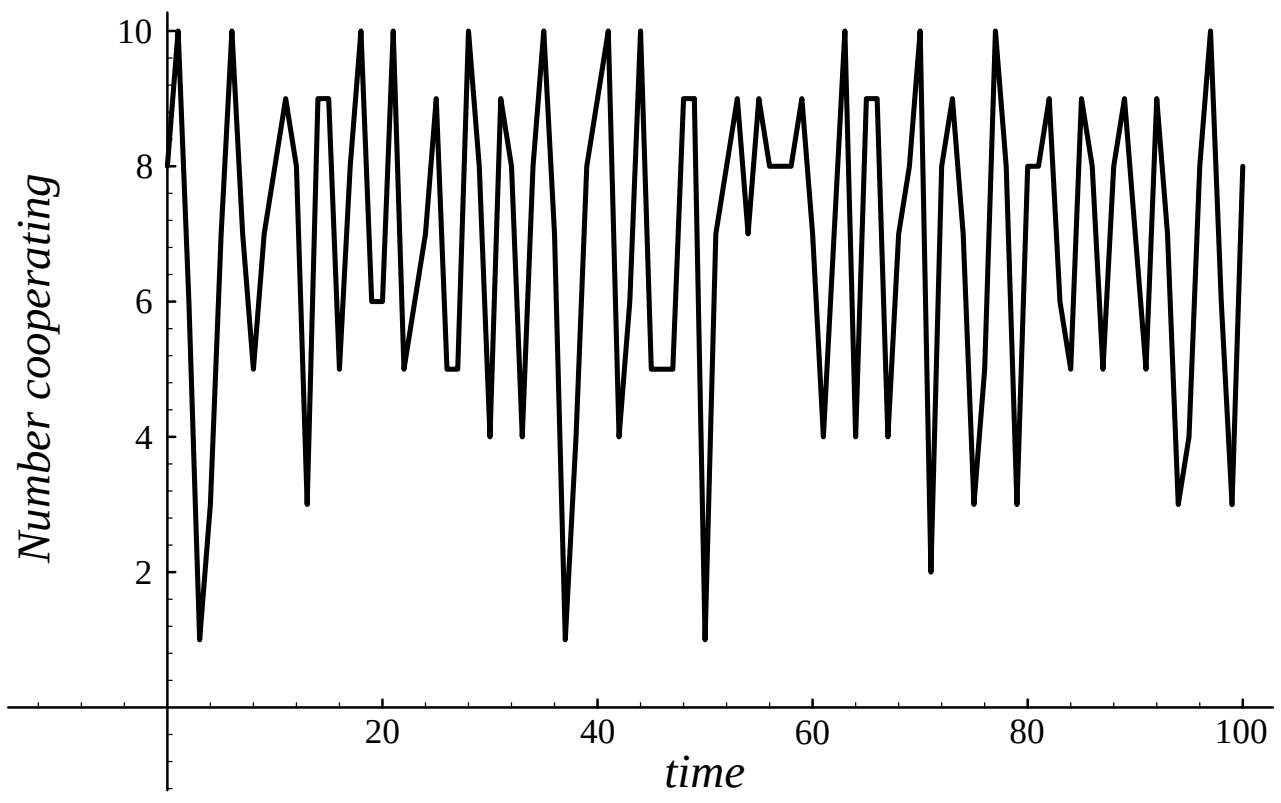

**Fig. 8.** Oscillatory dynamics of a group of opportunistic agents of size $n = 10$. The agents have horizon length $H = 10$, with $p = 0.99$, reevaluation rate $\alpha = 1$, information delay $\tau = 1$, and the same benefit and cost for cooperation as before.

As the uncertainty rises (which corresponds to decreasing $p$ in the experiments), the amplitude of the fluctuations decreases, until, for high enough uncertainty, the equilibrium point becomes stable and stochastic fluctuations dominate. The distinction between stochastic fluctuations and instability is an important one, and brings up possibility that one might be able to distinguish between randomness and deterministic chaos.

## *Bursty chaos*

If we decrease the horizon length of the opportunistic agents, we compromise the optimality of the mixed equilibrium in favor of that of the defecting one (i.e., the left minimum of the $\Omega$ function shifts downwards while the right one shifts upwards). Computer experiments show that for the shorter horizon length of $H = 5.5$, for example, the dynamics now flip-flops between two different behaviors, as can be seen in Fig. 9 (with $\tau = 4$ and with greater uncertainty $p = 0.9$ which speeds up the flip-flops). For long times, the system remains trapped in either the equilibrium point at $f_c = 0$, with only small fluctuations away, or in a chaotic wandering state. The latter behavior implies sensitivity to initial conditions even in the limit of small uncertainty, which makes long term predictions about the group's behavior impossible. Occasionally, the system breaks away from one of these two attractors only to be drawn back into the other, resulting in a bursty type of dynamics. Temporal averages are consequently a misleading indicator in such situations since bursty behavior implies that typical is not average for this kind of dynamics: given a brief snapshot, we cannot envision the longer history.

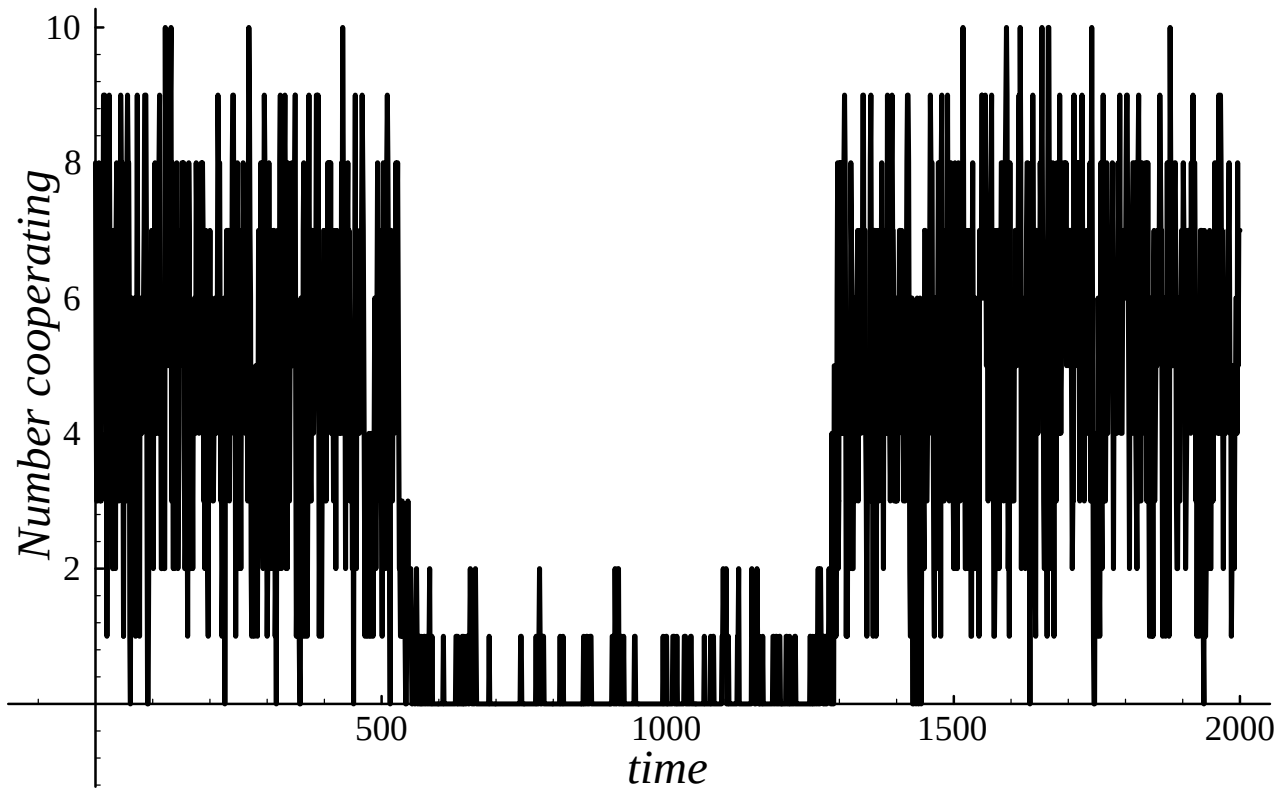

**Fig. 9.** Bursty stable and chaotic dynamics of a group of opportunistic agents of size $n = 10$. The agents have horizon length $H = 5.5$, with $p = 0.9$, reevaluation rate $\alpha = 1$, information delay $\tau = 4$, and the same benefit and cost for cooperation as before. The flip-flops between the two attractors of chaotic oscillations and the fixed point at overall defection extend into time beyond the time series shown here.

## 6. Conclusion

In this paper we have extended our previous study of the dynamics of social dilemmas to encompass a wide range of different individual beliefs. Within our framework of expectations all individuals are assumed to believe that a cooperative action on their part encourages further cooperation while defection encourages further defection. The unspecified part of the framework is how strong do individuals believe their influence to be. Originally, we choose bandwagon expectations as a simple way to capture the individual belief that the group will imitate one's actions to the extent that it is already behaving similarly. However, this runs counter to many people's intuition (thanks to self-reflection?) that when the rest of the group cooperates, people are very likely to free ride, expecting to have very little effect. The opportunistic class of expectations was introduced to capture this intuition and in the hope of anchoring our results in a more general set of expectations.

We attempted to map out the different types of beliefs using five classes of expectation functions whose functional forms are to be considered loosely as characterizing the classes rather than constraining them. We explored the two classes of bandwagon and opportunistic expectations in greater detail, since they seemed most realistic and interesting. We derived two general results for both types: (1) that there is an upper limit to the group size beyond which cooperation cannot be sustained; and (2) that, as long as the delays in information are small, there is a range of group sizes for which there are two fixed points for the dynamics, and that there can be sudden and abrupt transitions from the metastable equilibrium to the overall equilibrium if the group is observed for a long enough period of time.

The two fixed points for agents with bandwagon expectations are at either extremes of mutual defection and mutual cooperation, while for opportunistic agents, the second fixed point is mixed: a primarily cooperating group supports a defecting minority. For the case of opportunistic agents, we also found that delays in information caused the second, more cooperative fixed point to become unstable. As a result, the dynamics exhibits a panoply of behaviors, from opportunistic oscillations to bursty chaos, thus excluding the possibility of sustained cooperation over very long times.

There are clear implications that follow from this work for the possibility of cooperation in social groups and organizations. In order to achieve spontaneous cooperation over long periods of time, an organization made up of individuals with different beliefs and expectations should be structured into small subunits, with their members having access to timely information about the overall productivity of the system. This allows for the spontaneous emergence of cooperation and its persistence over long times. Failure to make available information about the overall utility accrued by the organization in a timely manner can lead to complicated patterns of unpredictable and unavoidable opportunistic defections, thus lowering the average level of cooperation in the system.